\documentclass[conference]{IEEEtran}
\ifCLASSINFOpdf
\else
\fi

\usepackage{cite}
\usepackage{amsmath,amssymb,amsfonts}
\usepackage{algorithmic}
\usepackage{graphicx}
\usepackage{textcomp}
\usepackage{xcolor}
\usepackage{hyperref}
\usepackage{graphicx}
\usepackage{subcaption}

\usepackage{tcolorbox}
\newcounter{remark}

\usepackage{braket}

\graphicspath{
{figure/}
}

\begin{document}

%

\title{Dependable Classical-Quantum \\Computer Systems Engineering}

\author{
author list
}


%

\author{\IEEEauthorblockN{
Edoardo Giusto\IEEEauthorrefmark{1},
Santiago Nuñez-Corrales\IEEEauthorrefmark{2,3},
Phuong Cao\IEEEauthorrefmark{2},\\
Alessandro Cilardo\IEEEauthorrefmark{1},
Ravishankar K. Iyer\IEEEauthorrefmark{3},
Weiwen Jiang\IEEEauthorrefmark{6},
Paolo Rech\IEEEauthorrefmark{4},
Flavio Vella\IEEEauthorrefmark{4},\\
Bartolomeo Montrucchio\IEEEauthorrefmark{5},
Samudra Dasgupta\IEEEauthorrefmark{7},
and Travis S. Humble \IEEEauthorrefmark{7}}
\IEEEauthorblockA{\IEEEauthorrefmark{1*}University of Naples Federico II, Italy - egiusto@ieee.org}
\IEEEauthorblockA{\IEEEauthorrefmark{2}National Center for Supercomputing Applications, Urbana, IL, US}
\IEEEauthorblockA{\IEEEauthorrefmark{3}University of Illinois Urbana-Champaign, US}
\IEEEauthorblockA{\IEEEauthorrefmark{4}University of Trento, Italy}
\IEEEauthorblockA{\IEEEauthorrefmark{5}
Politecnico di Torino, Italy}
\IEEEauthorblockA{\IEEEauthorrefmark{6}
George Mason University, VA, US}
\IEEEauthorblockA{\IEEEauthorrefmark{7}
Oak Ridge National Laboratory, TN, US}
}


\maketitle

\begin{abstract}
Quantum Computing (QC) offers the potential to enhance traditional High-Performance Computing (HPC) workloads by leveraging the unique properties of quantum computers, leading to the emergence of a new paradigm: HPC-QC.
While this integration presents new opportunities, it also brings novel challenges, particularly in ensuring the dependability of such hybrid systems. This paper aims to identify integration challenges, anticipate failures, and foster a diverse co-design for HPC-QC systems by bringing together QC, cloud computing, HPC, and network security. The focus of this emerging inter-disciplinary effort is to develop engineering principles that ensure the dependability of hybrid systems, aiming for a more prescriptive co-design cycle. Our framework will help to prevent design pitfalls and accelerate the maturation of the QC technology ecosystem. Key aspects include building resilient HPC-QC systems, analyzing the applicability of conventional techniques to the quantum domain, and exploring the complexity of scaling in such hybrid systems. This underscores the need for performance-reliability metrics specific to this new computational paradigm.
\end{abstract}


\IEEEpeerreviewmaketitle

\section{Introduction}
Exploiting the expected potential benefits Quantum Computing (QC) across scientific and engineering applications \cite{alexeev2023quantum} will require integrating QCs into standard HPC infrastructures, already the backbone of scientific computing.
This scenario ushers in a new computational paradigm: HPC-QC~\cite{
britt2017high
}.
Such paradigm entails the efficient and dependable incorporation of QPUs into standard HPC workflows \cite{saurabh2023conceptual,matsuura2022introducing,Tang2024Distributed}.
Scaling up emerging Noisy Intermediate-Scale Quantum (NISQ) devices and integrating them with HPC infrastructures pose significant engineering challenges.
Such classical-quantum integration has triggered exploration of multiple hardware \cite{britt2017high} and software \cite{saurabh2023conceptual} pathways starting from a model of Quantum Processing Units (QPUs) as remotely accessible hardware accelerators, akin to Graphics processing Units (GPUs), to much tighter, direct coupling within HPC systems. Recent advances highlight the integration of QPUs as HPC accelerators \cite{mccaskey2018language,mccaskey2020xacc}, the conceptualization of HPC-QPU enablers \cite{saurabh2023conceptual}, and the development of quantum kernels for scientific applications \cite{matsuura2022introducing}.

\begin{figure}[t]
    \centering
    \includegraphics[width=0.8\linewidth]{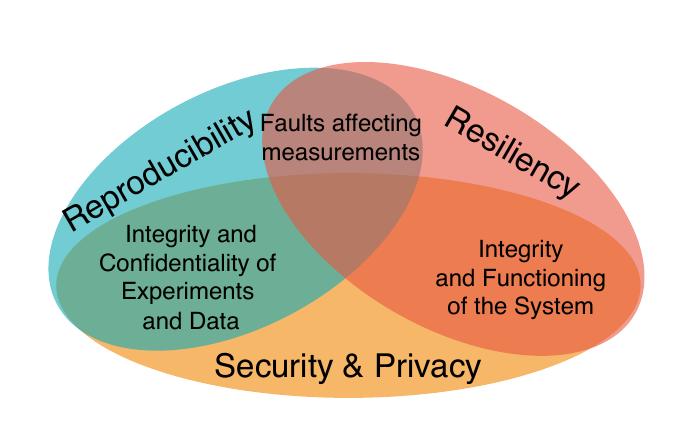}
    \caption{The Venn diagram highlights the intersection of reproducibility, resiliency and security in creating dependable computing systems. }
    \label{fig:dependability_venn}
\end{figure}

\begin{figure}[ht]
    \centering
    \includegraphics[width=0.55\linewidth]{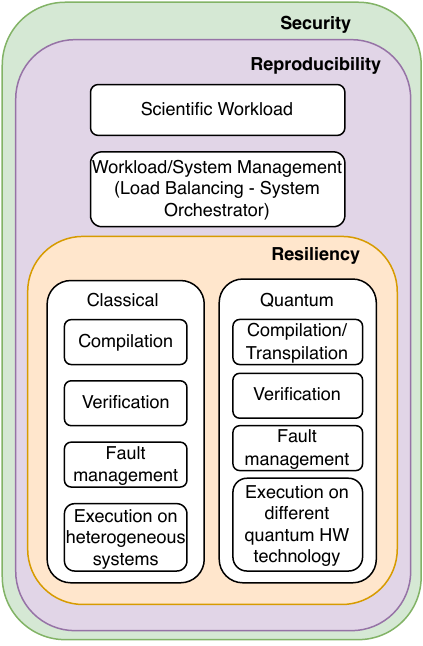}
    \caption{Dependability pillars in HPC-QC high-level architecture}
    \label{fig:high_level_architecture}
\end{figure}

In this paper, we highlight the need for dependable HPC-QC systems engineering by describing how aspects of dependability usually found in HPC infrastructure will apply to quantum platforms as their integration to classical resources deepens.
We do so by identifying how  \textit{reproducibility}, \emph{resiliency}, and \textit{security \& privacy} (Fig.~ \ref{fig:dependability_venn}) fit the new HPC-QC paradigm. These three pillars make up the definition of \textit{dependability} of a system.
The complexity derived from the intersection across pillars is non-trivial in emerging HPC-QC environments, and calls for an all-encompassing solution from a high-level view of computing architecture (Figure \ref{fig:high_level_architecture}).

Resiliency concerns both hardware and software at a very low abstraction level, concerning the compilation, verification, and execution of code on heterogeneous systems. A layer above we find reproducibility, managing the execution of a scientific calculation workflow in such a way that the produced output is consistent and repeatable between different executions and systems. Security eventually encompasses the whole spectrum of the system components, ensuring that the aforementioned workflow is not hampered by an external agent and that the processed data are untampered with and kept private.

The rest of the paper is organized in the following way: Sections \ref{sec:reproducibility}, \ref{sec:resiliency} and \ref{sec:security_privacy} address each of the pillars of dependability, while Section \ref{sec:discussion} brings them all together, laying out in the quest of dependable HPC-QC scale-up.
Eventually, Section \ref{sec:conclusions} wraps up the discussion.

\section{Reproducibility}
\label{sec:reproducibility}

Reproducibility is foundational to science and the ability to communicating and transferring knowledge reliably. In the presence of noise, quantum computers may become unreliable, in a formal sense, as their behaviors are no longer strictly predictable.  Statistical notions of operations and their outcomes become necessary for describing such noisy quantum computers. Moreover, results from quantum programs executed on noisy devices raise concerns for how to statistically quantify reproducibility in the presence of noise and errors
~\cite{dasgupta2022characterizing,dasgupta2022assessing,hu2023toward,senapati2023towards}.

\subsection{Accuracy vs. stability}

The \textit{accuracy} of a QC system represents the agreement between the observed and expected results of the QC program. For conventional analysis, binary outcomes from measuring the quantum register after execution of a quantum program (currently expressed as a quantum circuit) constitute the objects of interest. Histograms drawn from these outcomes characterize the \textit{stability} of the computation, which varies due to noise sources unpredictably due to non-stationary stochastic processes and their statistical observables \cite{dasgupta2023reliability}.
Whereas accuracy is a measure of the error in the calculation itself, stability quantifies the fluctuations in such observations with respect to time. Large and unpredictable fluctuations in these are a fundamental concern for the reproducibility of results. 

These uncertainties arise since practical efforts to build quantum computers introduce unexpected sources of noise through various types of imperfections \cite{gao2021practical,
wintersperger2023neutral}. In consequence, imperfections cause quantum devices to depart from idealized computing behavior. Example noise sources include spontaneous decay of qubits, leakage from the computational subspace, undesired external coupling due to spurious charge and magnetic fields, as well as inter-qubit cross-talk from capacitive coupling.  Similarly, noise in the control system arises from imperfections in the fundamental gate operations, e.g., in superconducting qubits, distortion and drift in microwave pulses often lead to errors. Externally, a multitude of mechanisms are put in place to isolate the quantum device and stave off decoherence as it interacts its environment (e.g., ultra-low temperatures, ultra-low vacuum). Thus anticipating how HPC environments will impinge on QC device operation becomes inescapable. 

\begin{tcolorbox}[width=\linewidth, colback=white!95!black, boxrule=0.5pt, left=2pt,right=2pt,top=1pt,bottom=1pt]
\stepcounter{remark}
{\bf Remark \arabic{remark}:}
{
The non-stationary noise in contemporary quantum devices presents a challenge for computational reproducibility that impacts the verification and validation of quantum computing demonstrations and hinders the production of trustworthy results.
}
\end{tcolorbox}

\subsection{Measuring quantum reproducibility}

To measure reproducibility in quantum computing systems, we extend the analysis of stability to discrete distributions with binary outcomes $\{f_b\}$ from quantum stochastic processes. To achieve this, we choose the Hellinger distance among various statistical distance measures as it extends to both discrete and continuous distributions, thus satisfying the requirements of a distance metric for comparison. For distributions $f(b)$ and $g(b)$, the Hellinger distance $H(f,g) \in [0,1]$ \cite{lindsay1994efficiency} is defined as

\begin{equation}
    H(f, g) = \sqrt{1-B(f,q)}
\end{equation}

with Bhattacharyya coefficient

\begin{equation}
    B(f,g) = \sum_{b} { \sqrt{ f(b) g(b)} }
\end{equation}

Outcomes are $\delta$-reproducible with tolerance $\epsilon$ when

\begin{equation}
\text{Pr}(H(f,g) \leq \epsilon) \geq 1 - \delta
\end{equation}

For quantum programs, $f$ and $g$ are computed for a given quantum circuit using multiple shots. As an example, the minimum sample size $L_\text{min}$ required for reproducibility for the Bernstein-Vazirani algorithm can be shown to be a non-linear function of the confidence level $1 - \delta$ and the accuracy threshold $\epsilon$:

\begin{equation}
L_\text{min} = z_\delta^2 \frac{p_r^{-2}-1}{p_r^{-2}(1-\epsilon)^2-1}
\end{equation}

where $z$ represents the standard normal variable (with mean 0 and variance 1), $z_{\delta}$ denotes the particular point where $\text{Pr}(z \geq z_{\delta}) = 1-\delta$, and $p_r$ signifies the probability of successfully identifying the secret string $r$ using the Bernstein-Vazirani algorithm in the presence of noise. A pressing concern is what values of the tolerances, $\epsilon$ and $\delta$, are sufficient for real-world applications of dependable QC. 

\section{Resiliency}
\label{sec:resiliency}
Resiliency characterizes the ability of a system to maintain a desired state given a range of perturbations, making it trusted and effective to accomplish its functionality, and capable of providing detection and graceful degradation of function and performance \cite{goerger2014engineered}. In computer systems, hardware and software components are expected to undertake this responsibility. At the software level, executable code must fulfill its purpose. To do so, \textit{compilation} of a quantum workload from circuits to pulses must be \textit{verified} to acquire \textit{trustworthiness} of compiled output.
At the hardware level, \textit{faults} arising in one or multiple physical units of the system should ideally not disrupt the execution flow or alter the reproducibility of results; more realistically, quantum program execution should be accompanied by information about the impact of faults. 
We briefly review here  compilation/transpilation and verification of quantum circuits, as well as faults in classical and quantum systems, and discuss both classical and quantum techniques to counteract such faults, analyzing whether it is possible to apply classical dependability techniques to the quantum domain.

\subsection{Compilation and verification}

Compiling (or transpiling) quantum circuits involves program transformations that match program constructs to quantum hardware capabilities and constraints. Constraints include native gate sets, qubit connectivity, coherence time, and qubit fidelity among others. This process seeks to optimize circuit performance while mitigating the effects of hardware limitations \cite{campbell2023superstaq,smith2019quantumISCA}. In tandem, quantum circuit verification is critical to ensure correctness and reliability. All this entails addressing both classical and quantum aspects derived from the probabilistic nature of quantum computation and noisy hardware. Existing methods include both simulation in classical computers and formal techniques which rigorously prove the correctness of the circuit \cite{wille2022verification}.

Compilation and verification of quantum programs must then be put in the context of hybrid HPC-QC environments. This brings a plethora of new concerns such as workload balancing optimization for specific and heterogeneous architectures (CPU+GPU+TPU+QPU), enabling parallel execution on such heterogeneous architectures, performance tuning, profiling and program optimization. Formal verification must then be overlaid across this complexity to guarantee that requirements are fulfilled, across the entire composite system.

\begin{tcolorbox}[width=\linewidth, colback=white!95!black, boxrule=0.5pt, left=2pt,right=2pt,top=1pt,bottom=1pt]
\stepcounter{remark}
{\bf Remark \arabic{remark}:}
{
Compilation processes must be adapted to heterogeneous HPC-QC architectures.
Quantum program execution must be verified across the entire stack, from high-level specifications to sequences of pulses.
}
\end{tcolorbox}

\subsection{Fault management}

\begin{figure}[t]
    \centering
    \includegraphics[width=0.5\textwidth]{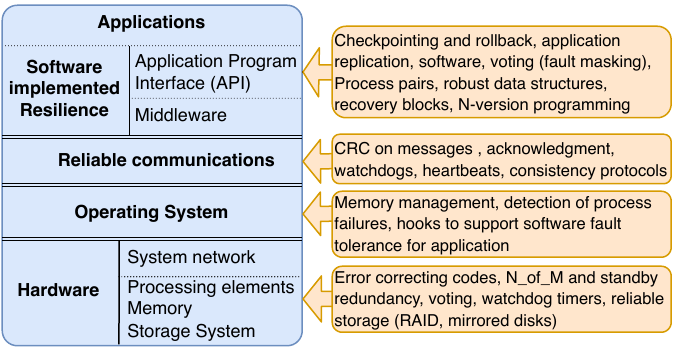}
    
    \caption{Classical techniques available at each level of a system.}
    \label{fig:dependability_techniques}
\end{figure}

Classical-quantum programs may fail to run for reasons well beyond compilation and verification errors. More generally, system or application failure can be caused by faults/errors at different levels of the system hierarchy (e.g., hardware, operating system, communication layer, middleware, or application). Their cascading effects thus call for a system perspective: we need to anticipate and engineer mechanisms that produce resilient computing systems as a result. Figure \ref{fig:dependability_techniques} poses relevant questions across the system hierarchy prior to building a resilient computing system \cite{dependable-computing-iyer}.

\begin{figure}[t]
    \centering
    \includegraphics[width=\linewidth]{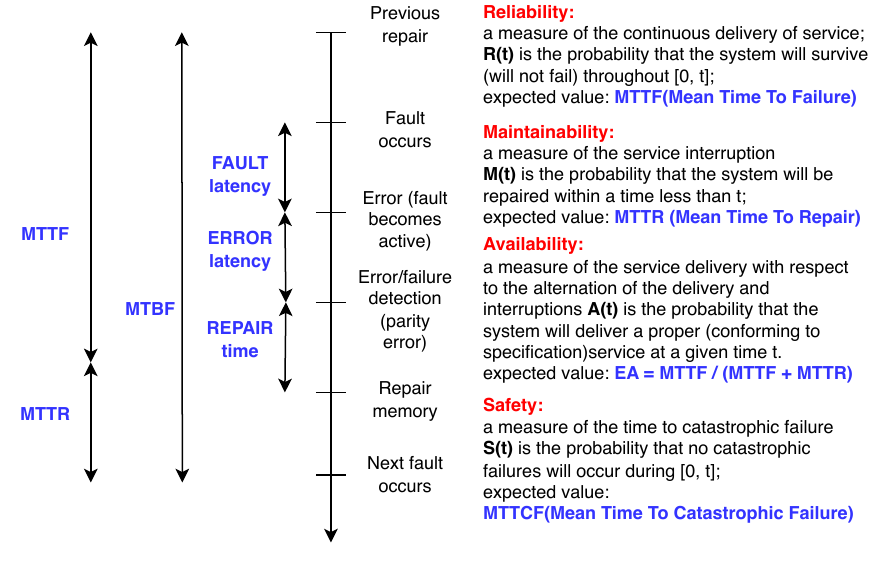}
    \caption{Resiliency measures in relation to the fault cycle of a classical system.}
    \label{fig:dependability}
\end{figure}

As with reproducibility, we need metrics to quantify the resiliency of an HPC-QC system.
We draw from prior research~\cite{dependable-computing-iyer} and suggest four metrics for the purpose at hand.
These metrics are quantified through mean time to failure (MTTF) or mean time to error (MTTE), mean time between failures (MTBF), mean time to catastrophic failure (MTTCF), fault latency, error latency, and repair time.
These metrics determine the extent of a \textit{fault cycle} (Figure~\ref{fig:dependability}).

\begin{itemize}
    \item \textbf{Reliability:} continuity of service delivery in the presence of hazards.
    \item \textbf{Maintainability:} the probability that the system will be repaired within a time less than $t$.
    \item \textbf{Availability:} service delivery for the alternation of the delivery and interruptions.
    \item \textbf{Safety:} time to catastrophic failure and unsafe system states.
\end{itemize}

In classical computing systems, particle impacts constitute a major source of faults. These are a naturally occurring byproduct of cosmic ray decay which corrupts stored values and the executed operations in both classical and quantum devices.  In a classical CMOS transistor, ionizing particles generate electron-hole pairs, releasing and depositing charge.
A sufficiently large deposited charge forces a transistor state to flip with three possible consequences. There may not be an effect on
program output if the fault is masked, or the corrupted data is not used. In another case, known as \textit{silent data corruption} (SDC), the program yields incorrect results yet continues to run. Finally, a bit flip may trigger a \textit{detected unrecoverable error} (DUE) in which the program crashes or the device is forced to reboot.

In QC devices, the impact of particles alters the state of qubit(s) by forcing them into decoherence. As an example, a fault mechanism in superconducting devices involves the generation of electron-hole pairs in the silicon substrate of the quantum chip, which in turn break Cooper pairs in the Josephson junction forming quasiparticles, that rapidly give rise to long-lasting phonons responsible for spreading the energy across the lattice of the quantum computer's substrate and interconnections \cite{nature_rad, Wilen2021}. While in a classical transistor the state is temporary reversed only if the deposited charge by the particle is higher than a threshold, in a qubit, even a single Cooper pair break is sufficient to disturb the quantum equilibrium, thus modifying the logic status.

Field experiments performed by Google AI on a 25 qubits array showed radiation-induced faults every tens of seconds \cite{mcewen2022resolving}. The reported error rate is several orders of magnitude higher than the one of modern CMOS technology. As a reference, the whole Titan supercomputer (composed of 14,000 nodes) has an error rate in the order of one error every few hours \cite{hpca2015}.
Finally, a \textit{Quantum Vulnerability Factor} has been recently proposed to quantify such effects at different levels of abstraction in the execution of quantum circuits \cite{Oliveira2023systematic,oliveira2022qufi}. At the physical level, \textit{hardware improvements} are continuously proposed to mitigate internal and external noise factors \cite{martinis2021saving,mcewen2024resisting}. 

\begin{tcolorbox}[width=\linewidth, colback=white!95!black, boxrule=0.5pt, left=2pt,right=2pt,top=1pt,bottom=1pt]
\stepcounter{remark}
{\bf Remark \arabic{remark}:}
{ Faults add randomness on top of intrinsic noise, changing the output probability distribution of the computation. Failures are likely to have larger impact on qubits than on classical CMOS transistors, and disrupt qubit behavior for a longer time.}
\end{tcolorbox}

HPC systems have led to the development of widespread techniques to handle faults (Fig. \ref{fig:dependability_techniques}). \textit{Checkpointing} involves periodically saving the state of a running computation to restart from in case of failure. While inapplicable to quantum states due to the exponential growth in information and inability to reconstruct arbitrary states, one can envision classical-like portions of quantum circuits where this may be possible. More generally, classical parts of quantum algorithms such as the variational quantum eigensolver (VQE) may be well suited for checkpointing.

\textit{Replication} instead consists of running the same on multiple processing units simultaneously. In QC, this is the main method used to estimate the resulting state vector's probabilities statistically. Finally, \textit{job resubmission} entails re-executing failed jobs, potentially on different resources, to improve the chances of successful completion. This can be done in QC if an error detection technique is implemented to identify sporadic external errors such as those induced by radiation, but not for errors due to inherent decoherence.

In the same direction, quantum error suppression (QES), quantum error mitigation (QEM) and quantum error correction (QEC) have drawn substantial interest. QES integrates resiliency directly into the hardware design, for instance by optimizing pulse engineering to minimize errors during gate operations \cite{mundada2023experimental}. QEM  addresses errors after their occurrence by statistically correcting noise effects on the final output \cite{cai2023quantum}. QEC employs redundancy across multiple physical qubits to detect and correct errors. This is not without challenges due to resource overhead and complex operations \cite{roffe2019quantum}, as well as the presence of correlated faults \cite{martinis2021saving}.

\begin{tcolorbox}[width=\linewidth, colback=white!95!black, boxrule=0.5pt, left=2pt,right=2pt,top=1pt,bottom=1pt]
\stepcounter{remark}
{\bf Remark \arabic{remark}:}
{
The Threshold and No-cloning Theorems place fundamental limitations on our ability to apply classical techniques to quantum computing systems. Classical-quantum systems will force the community to re-think resiliency across the stack.
}
\end{tcolorbox}

\section{Security \& Privacy}
\label{sec:security_privacy}
The complex interfaces between traditional HPC, QC, and emerging applications introduce hosts of new security issues:

\begin{itemize}
    \item \textbf{Challenge 1}: \textit{The security of classical systems is threatened Quantum-driven adversaries}, for example, traditional cryptography based on large integer factorization such as RSA can be broken by QC running Shor's algorithm.
    \item \textbf{Challenge 2}: \textit{The security of Quantum systems themselves}, for example, the confidentiality of information, the integrity of data, and the availability of Quantum Cloud services to the end user.
    \item \textbf{Challenge 3}: \textit{The privacy of user data and regulations when using HPC-QC to process sensitive research and health data} such as Controlled Unclassified Information (CUI) and Health Insurance Portability and Accountability Act (HIPAA) data.
\end{itemize}



We further lay out research directions for each of the challenging threat models in this hybrid domain as follows.

\subsection{Quantum-driven adversaries against Quantum and Classical systems}

With more powerful computation capabilities, attackers will leverage practical QPUs to upend traditional security assumptions, expected to break classical encryption in the next few decades.
In this context, the problem of adopting lattice-based cryptography is critically important.
Moreover, as GPUs become more powerful, we observe artificial intelligence-driven malware that automatically learns when and how to launch stealth data-stealing campaigns to minimize their footprints~\cite{10224946}.

\begin{tcolorbox}[width=\linewidth, colback=white!95!black, boxrule=0.5pt, left=2pt,right=2pt,top=1pt,bottom=1pt]
\stepcounter{remark}
{\bf Remark \arabic{remark}:}
{
Modern quantum-driven attacks will leverage the power of QPUs to \\ 1) mask the presence of attacks as natural faults \\
2) adapt attack traces to minimize their footprint, and \\
3) learn when, how, and where to launch attacks in a QC stack to maximize damage.
}
\end{tcolorbox}

\subsection{Attacks at the surface of HPC-QC integration}
New attack surfaces will emerge as physicists and computer scientists join forces to integrate HPC and QC, as both sides will make different assumptions on HPC-QC interfaces.
On the classical side, access control mechanisms such as federated authentication, key management, and computer network measurements must be adapted to the hybrid workload and data being transferred through the HPC-QC interface to gain visibility on adversaries.
As Quantum Key Distribution (QKD) systems and control mechanisms mature, the classical systems become the weakest link in HPC-QC integration and need careful security engineering and monitoring across the stack.
One of the possible approaches to security and performance issues in this context would be the development of a Quantum Internet able to interconnect the quantum part of HPC-QC systems.

\begin{tcolorbox}[width=\linewidth, colback=white!95!black, boxrule=0.5pt, left=2pt,right=2pt,top=1pt,bottom=1pt]
\stepcounter{remark}
{\bf Remark \arabic{remark}:}
{
Preemptive attack detection~\cite{cao2015preemptive} at the surface of HPC-QC is critical to ensure secure integration, which requires
\\ 1) inventing a new instrument that monitors quantum network data \\
2) distributing security monitoring instruments across networks for early detection of attacks \\
3) defining checkpointing and remedy mechanisms to respond to attacks.
}
\end{tcolorbox}





\subsection{Privacy concerns in integrated HPC-QC platforms}
Privacy-Perserving QC is critical as QC services, presumed to be untrusted, will be adopted in domains where data privacy and intellectual property are an inherent requirement,
such as drug discovery, financial optimization, and machine learning for health applications.
%
%
%

\subsubsection{Pure cryptography-based privacy-preserving approach}
Confidential QC approaches adopt classical cryptographic schemes for privacy-preserving computing, i.e. Fully Homomorphic Encryption (FHE) and Multi-Party Computation (MPC).
In particular, Blind Quantum Computing (BQC)~\cite{fitzsimons17}
partially delegates quantum computation to a remote server without disclosing the computation itself.
More recent proposals for BQC relaxed this requirement~\cite{huang17}, but still need an assumption of multiple non-colluding servers, which may not be realistic in practice.
Quantum Homomorphic Encryption (QHE)~\cite{ouyang15}, on the other hand,
enables the quantum computer to work on encrypted quantum data and produce an encrypted outcome, without accessing data, similar to its classical counterpart,
but it remains impractical as it involves exponential computational overheads.
As an interesting development, there are some attempts to come up with special-purpose cryptographic schemes for privacy-preserving computing,
i.e. schemes which can only be applied for specific types of computation.
For example, the authors of~\cite{ayanzadeh23} propose a privacy-preserving scheme specific to Quantum Approximate Optimization Algorithm,
which achieves realistic performance and proves to be feasible on current and near-term quantum computers.

%

\subsubsection{A Trusted Quantum Computing Base (TQCB) for privacy-preserving}

Classical Trusted Execution Environments (TEEs) rely on hardware that isolates computation within a chip, offering security. 
Quantum computers, with their distributed data and components, pose challenges to redefining Trusted Computing Base~\cite{trochatos23} as researchers are exploring ways to establish natively quantum TEEs for protecting sensitive data.

\begin{tcolorbox}[width=\linewidth, colback=white!95!black, boxrule=0.5pt, left=2pt,right=2pt,top=1pt,bottom=1pt]
\stepcounter{remark}
{\bf Remark \arabic{remark}:}
{
Standardizing data privacy and security specifications using mathematical formalisms, and more importantly making specifications \textit{executable}, is critically needed to enable:  \\ 1) secure-by-construction data access and sharing protocols, \\
2) formal verification of the entire quantum computing stacks and synthesizing corresponding correct implementations, and \\
3) testbed for evaluating different attack scenarios such as data leaks or compromising root of trust.
}
\end{tcolorbox}

We expect to engage with NIST to formally define the above standard needs.







\begin{figure*}[t]
    \centering
    \includegraphics[width=1\textwidth]{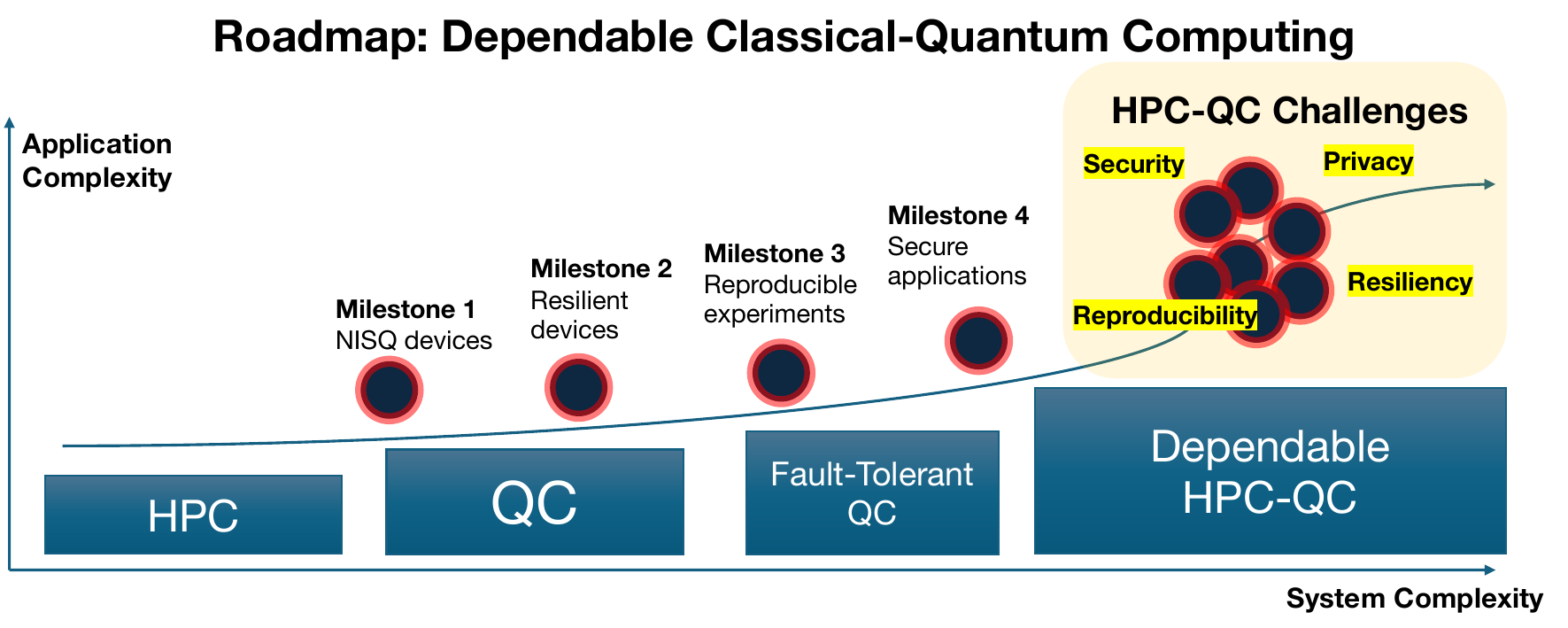}
    \caption{Roadmap for Dependable HPC-QC.
    }
    \label{fig:roadmap}
\end{figure*}

\section{Discussion}
\label{sec:discussion}
To become a dependable technology capable of delivering actual value to users with scientific applications in need of performance, quantum hardware platforms will need to adapt to the pressures and environments HPC systems already operate. The question of how to successfully perform this HPC-QC integration is far from answered since differences between classical and quantum technologies, and the state of maturity across both fields introduce additional confounding factors.

Our work --as well as cited works from others- suggests that a new area of research is needed at the intersection between classical HPC and quantum computing.
The title of this article, \textit{Dependable Classical-Quantum Computing Systems Engineering}, deliberately qualifies the nature of the endeavor as one drawing heavily from scientific principles to articulate usable systems, not to advance the science of devices \textit{per se}.
Experimental quantum testbeds dedicated to research have been instrumental in advancing the state of the art in quantum hardware; our views suggest we need actual HPC-QC testbeds dedicated to a holistic and rigorous interrogation of dependability.

The three pillars of dependability are deeply intertwined into HPC-QC systems: the more application and research users depend on these, the more sophisticated the ability of the system to satisfy essential guarantees must be.
Starting from the high-level definition of a scientific workload, application users are concerned with the solution to their problem --i.e. computing as a service-, with little to no practical interest in the lower levels of the stack that satisfy their needs.
However, application users do concern themselves with performance, reproducibility, resiliency, and security in their scientific work.
Building dependable classical-quantum hardware and software components multiplies the complexity of standard dependability concerns across HPC.

At a technical level, the ability to make progress across multiple directions requires transparent access to software and hardware systems.
HPC systems are currently mostly structured around open-source stacks where the code is fully inspectable, and the hardware can be easily interrogated at every meaningful level.
A similar ecosystem of tools and software for classical-quantum computing has yet to consolidate.
We do observe an increasing number of standard packages used in quantum hardware research and practice; ensuring these remain accessible and more are produced constitutes a substantial objective toward dependable HPC-QC.
In particular, reference implementations that are open, transparent, and accessible become highly desirable.

We believe it is necessary to fully engage the HPC, networking, and QC communities, with three main goals.
The first goal is to delimit specific problem areas where we only have intuitions of how to address them so far: the collection of remarks derived here points to joint, interdisciplinary work to address multiple sources of uncertainty.
The second goal is to identify milestones in the development of dependable HPC-QC systems that synergistically benefit from ongoing work by members of the QC community.
These milestones must be connected to progress in quantum hardware testbeds and accelerate hardware-software co-design of full systems.
A successful selection of milestones will be characterized by a healthy balance between new avenues of exploration and effective ways to discard unproductive research directions.
Third and last, we seek to foster a larger conversation resulting in a roadmap akin to that in Figure \ref{fig:roadmap} capable of fully realizing a long-term, sustainable HPC-QC integration program.

\section{Conclusion}
\label{sec:conclusions}

This article shows that, despite experience gathered across several decades of HPC practice, the introduction of QC brings new and interesting challenges. Ensuring the dependability of hybrid systems is paramount in order to leverage such computational power for scientific computing applications.
This paper describes the three pillars of dependability - reproducibility, resiliency, and security \& privacy - in the context of HPC-QC integration.
Reproducibility is impacted by quantum noise, while classical approaches in the resiliency domain may have fundamental shortcomings if applied to the quantum realm. Security threats are amplified due to the intricate integration of such heterogeneous components. 
Overcoming the dependability challenges is crucial for enabling reliable, high-performance scientific applications leveraging quantum resources.
This is a call to arms to experts across HPC, QC, and cybersecurity communities to come together to address these issues.
We argue that developing a combined co-design approach, supported by open testbeds, can pave the way towards dependable, robust HPC-QC platforms that can deliver on the revolutionary promise of this emerging paradigm.

\newpage
\clearpage
\bibliographystyle{unsrt}
\bibliography{refs}

\begin{thebibliography}{10}

\bibitem{alexeev2023quantum}
Yuri Alexeev, Maximilian Amsler, Paul Baity, Marco~Antonio Barroca, Sanzio Bassini, Torey Battelle, Daan Camps, David Casanova, Frederic~T Chong, Charles Chung, et~al.
\newblock Quantum-centric supercomputing for materials science: A perspective on challenges and future directions.
\newblock {\em arXiv preprint arXiv:2312.09733}, 2023.

\bibitem{britt2017high}
Keith~A Britt and Travis~S Humble.
\newblock High-performance computing with quantum processing units.
\newblock {\em ACM Journal on Emerging Technologies in Computing Systems (JETC)}, 13(3):1--13, 2017.

\bibitem{saurabh2023conceptual}
Nishant Saurabh, Shantenu Jha, and Andre Luckow.
\newblock A conceptual architecture for a quantum-hpc middleware.
\newblock In {\em 2023 IEEE International Conference on Quantum Software (QSW)}, pages 116--127. IEEE, 2023.

\bibitem{matsuura2022introducing}
AY~Matsuura and Timothy~G Mattson.
\newblock Introducing the quantum research kernels: Lessons from classical parallel computing.
\newblock {\em arXiv preprint arXiv:2211.00844}, 2022.

\bibitem{Tang2024Distributed}
Wei Tang and Margaret Martonosi.
\newblock Distributed quantum computing via integrating quantum and classical computing.
\newblock {\em Computer}, 57(4):131--136, 2024.

\bibitem{mccaskey2018language}
Alexander~J McCaskey, Eugene~F Dumitrescu, Dmitry Liakh, Mengsu Chen, Wu-chun Feng, and Travis~S Humble.
\newblock A language and hardware independent approach to quantum--classical computing.
\newblock {\em SoftwareX}, 7:245--254, 2018.

\bibitem{mccaskey2020xacc}
Alexander~J McCaskey, Dmitry~I Lyakh, Eugene~F Dumitrescu, Sarah~S Powers, and Travis~S Humble.
\newblock Xacc: a system-level software infrastructure for heterogeneous quantum--classical computing.
\newblock {\em Quantum Science and Technology}, 5(2):024002, 2020.

\bibitem{dasgupta2022characterizing}
Samudra Dasgupta and Travis~S Humble.
\newblock Characterizing the reproducibility of noisy quantum circuits.
\newblock {\em Entropy}, 24(2):244, 2022.

\bibitem{dasgupta2022assessing}
Samudra Dasgupta and Travis~S Humble.
\newblock Assessing the stability of noisy quantum computation.
\newblock In {\em Quantum Communications and Quantum Imaging XX}, volume 12238, pages 44--49. SPIE, 2022.

\bibitem{hu2023toward}
Zhirui Hu, Robert Wolle, Mingzhen Tian, Qiang Guan, Travis Humble, and Weiwen Jiang.
\newblock Toward consistent high-fidelity quantum learning on unstable devices via efficient in-situ calibration.
\newblock In {\em 2023 IEEE International Conference on Quantum Computing and Engineering (QCE)}, volume~1, pages 848--858. IEEE, 2023.

\bibitem{senapati2023towards}
Priyabrata Senapati, Zhepeng Wang, Weiwen Jiang, Travis~S Humble, Bo~Fang, Shuai Xu, and Qiang Guan.
\newblock Towards redefining the reproducibility in quantum computing: A data analysis approach on nisq devices.
\newblock In {\em 2023 IEEE International Conference on Quantum Computing and Engineering (QCE)}, volume~1, pages 468--474. IEEE, 2023.

\bibitem{dasgupta2023reliability}
Samudra Dasgupta and Travis~S Humble.
\newblock Reliability of noisy quantum computing devices.
\newblock {\em arXiv preprint arXiv:2307.06833}, 2023.

\bibitem{gao2021practical}
Yvonne~Y Gao, M~Adriaan Rol, Steven Touzard, and Chen Wang.
\newblock Practical guide for building superconducting quantum devices.
\newblock {\em PRX Quantum}, 2(4):040202, 2021.

\bibitem{wintersperger2023neutral}
Karen Wintersperger, Florian Dommert, Thomas Ehmer, Andrey Hoursanov, Johannes Klepsch, Wolfgang Mauerer, Georg Reuber, Thomas Strohm, Ming Yin, and Sebastian Luber.
\newblock Neutral atom quantum computing hardware: performance and end-user perspective.
\newblock {\em EPJ Quantum Technology}, 10(1):32, 2023.

\bibitem{lindsay1994efficiency}
Bruce~G Lindsay.
\newblock Efficiency versus robustness: the case for minimum hellinger distance and related methods.
\newblock {\em The annals of statistics}, 22(2):1081--1114, 1994.

\bibitem{goerger2014engineered}
Simon~R Goerger, Azad~M Madni, and Owen~J Eslinger.
\newblock Engineered resilient systems: A dod perspective.
\newblock {\em Procedia Computer Science}, 28:865--872, 2014.

\bibitem{campbell2023superstaq}
Colin Campbell, Frederic~T Chong, Denny Dahl, Paige Frederick, Palash Goiporia, Pranav Gokhale, Benjamin Hall, Salahedeen Issa, Eric Jones, Stephanie Lee, et~al.
\newblock Superstaq: Deep optimization of quantum programs.
\newblock In {\em 2023 IEEE International Conference on Quantum Computing and Engineering (QCE)}, volume~1, pages 1020--1032. IEEE, 2023.

\bibitem{smith2019quantumISCA}
Kaitlin~N Smith and Mitchell~A Thornton.
\newblock A quantum computational compiler and design tool for technology-specific targets.
\newblock In {\em Proceedings of the 46th International Symposium on Computer Architecture}, pages 579--588, 2019.

\bibitem{wille2022verification}
Robert Wille and Lukas Burgholzer.
\newblock Verification of quantum circuits.
\newblock {\em Handbook of Computer Architecture}, pages 1--28, 2022.

\bibitem{dependable-computing-iyer}
Ravishankar K.~Iyer Zbigniew T.~Kalbarczyk, Nithin M.~Nakka.
\newblock {\em \\Dependable Computing: Design and Assessment}, chapter~1.
\newblock \\ISBN: 978-1-118-70944-3, John Wiley \& Sons, 2024.

\bibitem{nature_rad}
Antti~P. Veps{\"a}l{\"a}inen, Amir~H. Karamlou, John~L. Orrell, Akshunna~S. Dogra, Ben Loer, Francisca Vasconcelos, David~K. Kim, Alexander~J. Melville, Bethany~M. Niedzielski, Jonilyn~L. Yoder, Simon Gustavsson, Joseph~A. Formaggio, Brent~A. VanDevender, and William~D. Oliver.
\newblock Impact of ionizing radiation on superconducting qubit coherence.
\newblock {\em Nature}, 584(7822):551--556, 2020.

\bibitem{Wilen2021}
C.~D. Wilen, S.~Abdullah, N.~A. Kurinsky, C.~Stanford, L.~Cardani, G.~D'Imperio, C.~Tomei, L.~Faoro, L.~B. Ioffe, C.~H. Liu, A.~Opremcak, B.~G. Christensen, J.~L. DuBois, and R.~McDermott.
\newblock Correlated charge noise and relaxation errors in superconducting qubits.
\newblock {\em Nature}, 594(7863):369--373, 2021.

\bibitem{mcewen2022resolving}
Matt McEwen, Lara Faoro, Kunal Arya, Andrew Dunsworth, Trent Huang, Seon Kim, Brian Burkett, Austin Fowler, Frank Arute, Joseph~C Bardin, et~al.
\newblock Resolving catastrophic error bursts from cosmic raysrin large arrays of superconducting qubits.
\newblock {\em Nature Physics}, 18(1):107--111, 2022.

\bibitem{hpca2015}
D.~{Tiwari}, S.~{Gupta}, J.~{Rogers}, D.~{Maxwell}, P.~{Rech}, S.~{Vazhkudai}, D.~{Oliveira}, D.~{Londo}, N.~{DeBardeleben}, P.~{Navaux}, L.~{Carro}, and A.~{Bland}.
\newblock Understanding {GPU} errors on large-scale {HPC} systems and the implications for system design and operation.
\newblock In {\em 2015 IEEE 21st International Symposium on High Performance Computer Architecture (HPCA)}, pages 331--342, 2015.

\bibitem{Oliveira2023systematic}
Daniel Oliveira, Edoardo Giusto, Betis Baheri, Qiang Guan, Bartolomeo Montrucchio, and Paolo Rech.
\newblock A systematic methodology to compute the quantum vulnerability factors for quantum circuits.
\newblock {\em IEEE Transactions on Dependable and Secure Computing}, pages 1--15, 2023.

\bibitem{oliveira2022qufi}
Daniel Oliveira, Edoardo Giusto, Emanuele Dri, Nadir Casciola, Betis Baheri, Qiang Guan, Bartolomeo Montrucchio, and Paolo Rech.
\newblock Qufi: a quantum fault injector to measure the reliability of qubits and quantum circuits.
\newblock In {\em 2022 52nd Annual IEEE/IFIP International Conference on Dependable Systems and Networks (DSN)}, pages 137--149. IEEE, 2022.

\bibitem{martinis2021saving}
John~M Martinis.
\newblock Saving superconducting quantum processors from decay and correlated errors generated by gamma and cosmic rays.
\newblock {\em npj Quantum Information}, 7(1):90, 2021.

\bibitem{mcewen2024resisting}
Matt McEwen, Kevin~C Miao, Juan Atalaya, Alex Bilmes, Alex Crook, Jenna Bovaird, John~Mark Kreikebaum, Nicholas Zobrist, Evan Jeffrey, Bicheng Ying, et~al.
\newblock Resisting high-energy impact events through gap engineering in superconducting qubit arrays.
\newblock {\em arXiv preprint arXiv:2402.15644}, 2024.

\bibitem{mundada2023experimental}
Pranav~S Mundada, Aaron Barbosa, Smarak Maity, Yulun Wang, Thomas Merkh, TM~Stace, Felicity Nielson, Andre~RR Carvalho, Michael Hush, Michael~J Biercuk, et~al.
\newblock Experimental benchmarking of an automated deterministic error-suppression workflow for quantum algorithms.
\newblock {\em Physical Review Applied}, 20(2):024034, 2023.

\bibitem{cai2023quantum}
Zhenyu Cai, Ryan Babbush, Simon~C Benjamin, Suguru Endo, William~J Huggins, Ying Li, Jarrod~R McClean, and Thomas~E O’Brien.
\newblock Quantum error mitigation.
\newblock {\em Reviews of Modern Physics}, 95(4):045005, 2023.

\bibitem{roffe2019quantum}
Joschka Roffe.
\newblock Quantum error correction: an introductory guide.
\newblock {\em Contemporary Physics}, 60(3):226--245, 2019.

\bibitem{10224946}
Keywhan Chung, Phuong Cao, Zbigniew~T. Kalbarczyk, and Ravishankar~K. Iyer.
\newblock stealthml: Data-driven malware for stealthy data exfiltration.
\newblock In {\em 2023 IEEE International Conference on Cyber Security and Resilience (CSR)}, pages 16--21, 2023.

\bibitem{cao2015preemptive}
Phuong Cao, Eric Badger, Zbigniew Kalbarczyk, Ravishankar Iyer, and Adam Slagell.
\newblock Preemptive intrusion detection: Theoretical framework and real-world measurements.
\newblock In {\em Proceedings of the 2015 Symposium and Bootcamp on the Science of Security}, pages 1--12, 2015.

\bibitem{fitzsimons17}
Joseph~F. Fitzsimons.
\newblock Private quantum computation: an introduction to blind quantum computing and related protocols.
\newblock {\em npj Quantum Information}, 3(1):23, Jun 2017.

\bibitem{huang17}
He-Liang Huang, Qi~Zhao, Xiongfeng Ma, Chang Liu, Zu-En Su, Xi-Lin Wang, Li~Li, Nai-Le Liu, Barry~C. Sanders, Chao-Yang Lu, and Jian-Wei Pan.
\newblock Experimental blind quantum computing for a classical client.
\newblock {\em Phys. Rev. Lett.}, 119:050503, Aug 2017.

\bibitem{ouyang15}
Yingkai Ouyang, Si-Hui Tan, and Joseph Fitzsimons.
\newblock Quantum homomorphic encryption from quantum codes.
\newblock {\em Physical Review A}, 2015.

\bibitem{ayanzadeh23}
Ramin Ayanzadeh, Ahmad Mousavi, Narges Alavisamani, and Moinuddin Qureshi.
\newblock Enigma: Privacy-preserving execution of qaoa on untrusted quantum computers, 2023.

\bibitem{trochatos23}
T.~Trochatos, C.~Xu, S.~Deshpande, Y.~Lu, Y.~Ding, and J.~Szefer.
\newblock A quantum computer trusted execution environment.
\newblock {\em IEEE Computer Architecture Letters}, 22(02):177--180, jul 2023.

\end{thebibliography}

\end{document}